\documentclass{aa}  
\usepackage{graphicx}
\usepackage{txfonts}
\begin{document}

   \title{Planetary transit timing variations induced by stellar binarity}

   \subtitle{The light travel time effect}

   \author{M. Montalto
          \inst{1,2}
           }

   \offprints{M. Montalto}

   \institute{
              Max-Planck-Institute for Extraterrestrial Physics, Giessenbachstr.,
              Garching b Muenchen, 85741, Germany.\\
    \and
              Universitaetssternwarte  Muenchen,   Scheiner  Str. 1,  D 81679
              Muenchen, Germany.
             \email{montalto@usm.uni-muenchen.de}
   }

   \date{}

 
  \abstract
   {
    Since the discovery of the first transiting extrasolar planet, transit timing has been recognized as
    a powerful method to discover and characterize additional planets in these systems. 
    However, the gravitational influence of additional
    planets is not the only expected source of transit timing
    variations. 
   }
   {
    In this work, we derive the expected detection frequency of stellar companions
    of hot-jupiter transiting planets host-stars, detectable by means of transit timing analysis.
    Since roughly half of the stars in the solar neighborhood belong to binary
    or multiple stellar systems, the same fraction of binary systems may be expected to be present
    among transiting planet-host stars, unless planet formation is significantly
    influenced by the presence of a stellar companion. Transit searches are less affected by the
    selection biases against long-period binaries that plague radial velocity surveys. 
   }
   {
    We considered the frequency and the period, mass ratio and eccentricity distributions of 
    known binary systems in the solar neighborhood, and estimated the fraction of transiting planet-hosts
    expected to show detectable transit timing variations due to the light travel time effect
    in a binary stellar system, in function of the time since the discovery of the planet.
   }
   {
    If the frequency of binaries among hot-jupiter planet host stars is the same as determined in the solar neighborhood,
    after 5 years since the discovery of a sample of transiting planets $1.0\%\pm0.2\%$ of them have a probability $>99\%$ to present 
    transit timing variations $>50$ sec induced by stellar binarity, 
    and $2.8\%\pm0.3\%$ after 10 years, if the planetary and binary orbits are coplanar. Considering the case 
    of random inclinations the probabilities are $0.6\%\pm0.1\%$ and $1.7\%\pm0.2\%$ after 5 and 10 years respectively.
    Our estimates can be considered conservative lower limits, since we have taken into account only binaries
    with periods $P>5\,\cdot\,10^3$ days ($a\gtrsim6$ AU). Our simulations indicate that transit timing variations due
    to the light travel time effect will allow us to discover stellar companions up to maximum separations equal to $a\sim36$ AU
    after 5 years since the discovery of the planet ($a\sim75$ AU after 10 years). 
   }
   {
    Comparing the results of the observations with the above predictions allows to understand if stellar companions
    at critical separations ($<100$ AU) are favoring or hindering the formation of hot-jupiter planets. 
    Comparing the results of transit timing detections with those obtained by other complementary methods
    results in a more complete determination of stellar multiplicity around transiting planet-host 
    stars. Moreover, transit timing analysis allows us to probe stellar multiplicity at critical separations ($<100$ AU) around
    stars located in different regions of the Galaxy, and not just in the solar neighborhood.    
   }

   \keywords{planetary systems -- techniques: photometric -- methods: numerical, occultations}

   \maketitle

%

\section{Introduction}
\label{s:introduction}

The frequency of binary stars and multiple stellar systems around solar-like stars
in the solar neighborhood has been extensively studied in the past.
Duquennoy \& Mayor~(1991), from a
sample of 164 primary G-dwarf stars analyzed during almost 13 yr with the 
CORAVEL spectrograph, obtained that the ratios of single:double:triple:quadruple
systems are 51:40:7:2 respectively. 
This fraction of binaries should be
present also in a sample of planet host stars if planets do form in any kind of binary systems
and there are no selection biases applied to define the sample of stars 
where the planets are searched for.

From a theoretical point of view the problem of the formation of giant planets in close
binary systems is still largely debated. Truncation and heating of circumstellar protoplanetary
disks are expected to occur in close binary systems
(Artymowics \& Lubow~1994; Nelson~2000). However there is no general consensus on which 
consequences these processes have on planet formation. Nelson~(2000) 
showed that the formation of giant planets is unlikely in equal mass binaries with semi-major
axis $\sim50$ AU, both by means of the disk instability (Boss~1997) and the
core accretion mechanisms (Pollack et al.~1996). On the contrary Boss~(2006), showed that
in the context of the disk instability scenario, the presence of a close-by stellar companion 
may in fact trigger clumps formation leading to giant planets. Along the same lines
Duch\^{e}ne (2010) suggested that in tight binaries ($a<100$ AU) massive planets
are formed by disk instability at the same rate as less massive 
planets in wider binaries and single stars.

According to Marzari et al.~(2005), independently from 
the planet formation mechanism, tidal perturbations of the companion star influence both 
the onset of instability and the following chaotic evolution of protoplanetary disks. 
In particular several studies have pointed out that the gravitational influence of stellar companions 
on the dynamics of planetary systems becomes significant at separations $\le100$ AU
(e.g. Pfhal \& Muterspaugh~2006; Desidera \& Barbieri~2007; Duch\^{e}ne~2010). 
Despite that, four planetary systems have been discovered in binaries with 
separations around $20$ AU: Gamma Cephei
(Hatzes et al.~2003), Gliese 86 (Queloz et al.~2000; Els et al.~2001), HD 41004 (Santos et al.~2002),
and HD196885 (Correia et al.~2008).
These observational results are clearly challenging our current knowledge of planet formation
and evolution in binary systems.

Several imaging surveys have successfully identified stellar companions
to planet-host stars (e.g. Patience et al.~2002; 
Luhman \& Jayawardhana~2002; Chauvin et al.~2006; Mugrauer et al.~2007; 
Eggenberger et al.~2007; Eggenberger \& Udry 2007). At present, giant planets
around wide binary systems appear as frequent as planets around single stars
(Bonavita \& Desidera~2007), suggesting that wide binaries are not significantly altering planet
formation processes. 
However, there is a marginal statistical evidence that binaries with separations smaller than
100 AU may have a lower frequency of planetary systems than around single stars.
Eggenberger et al.~(2008) estimated a difference in the binary frequency between carefully
selected samples of non planet-host and planet-host stars ranging between $8.2\%\,\pm\,5.0\%$
and $12.5\%\,\pm\,5.9\%$, considering binaries with semi-major axis between 35 AU and 250 AU,
further pointing out that this difference seems mostly evident for binaries with $a<100$ AU.
These results are obtained using samples of stars hosting radial velocity discovered planets, 
for the obvious reason that Doppler spectroscopy
has been so far the most successful planet detection method, providing then the largest
sample of planets from which statistical conclusions can be drawn. Nevertheless,
planets discovered with this technique are known to be adversely selected against
close binaries. Doppler spectroscopy is complicated by light contamination
of stellar companions, and blending of spectral lines. This implies that radial
velocity surveys typically exclude known moderately and close binaries from their target lists.
As a consequence the occurrence of planetary systems in binaries, and in particular at 
critical separations ($\le100$ AU), is still poorly constrained by the observations.
While the use of a control sample of stars, proposed by Eggenberger et al.~(2007, 2008), is expected
to mitigate the impact on the results given by the bias against close binaries applied by radial velocity surveys,
other samples of planet-host stars and different techniques may be useful to probe the
frequency of planets in binary stellar systems. 

While radial velocity planet searches tend to exclude binaries from their target lists, transit searches do not 
apply $a\,priori$ selection criteria against binaries. 
Transiting planets are routinely searched among all stars, and then subjected to 
follow-up photometric and spectroscopic analysis. 
Photometric analysis aims mainly at ruling out grazing eclipsing binary stellar systems
which manifest themselves by means of markedly V-shaped eclipses, the presence of secondary eclipses, 
color changes during the eclipses and 
light modulations with the same periodicity of the transiting object.
Spectroscopic analysis is then used to further rule-out giant stars
primaries and the more complicated scenarios involving hierarchical triple systems with an eclipsing binary
stellar system, and blends with background eclipsing binaries (Brown 2003). However, these
follow-up analysis do not eliminate planets in binary stellar systems.
Several transiting planets are already known members of binaries (e. g. Daemgen et al.~2009),
and others have suspected close companions as indicated by the presence of radial velocity
and transit timing variations (Winn et al.~2010; Maxted et al.~2010; Queloz et al.~2010; Rabus et al.~2009). 
Moreover for transiting planets we have a firm constraint on the planetary orbital inclination (which is close to $90^{\circ}$).
The Rossiter-McLaughlin effect (Rossiter 1924; McLaughlin 1924) can be used 
to probe the sky-projected angle $\beta$ between the stellar rotation axis and the planet's orbital axis.
By transforming the projected angle $\beta$ into the the real spin-orbit angle $\psi$ using a statistical
approach and the entire sample of planets with Rossiter-McLaughlin measurements, Triaud et al.~(2010)
derived that most transiting planets have misaligned orbits (80$\%$ with $\psi>22^{\circ}$), and that
the histogram of projected obliquities closely reproduces the theoretical distribution of $\psi$
using Kozai cycles and tidal friction (Fabrycky \& Tremaine 2007). Since type I and II migration are not able to explain
the present observations, the indication is that the Kozai mechanism is the major responsible of the formation
of hot-jupiter planets. In this case, we should expect that most hot-jupiter planets
have stellar companions. The discovery of close stellar companions to transiting planets systems
is then of primary importance, since it would constitute a strong proof in favor of the Kozay cycles and tidal friction mechanism.

Transiting planets host stars then constitute an interesting sample of objects where to look for additional distant companions. 
The presence of a stellar companion around these stars can be inferred at least by means of four independent and complementary techniques: 
transit timing variations, radial velocity drifts, direct imaging and IR excess. 

Deriving the expected $frequency$ of transiting planet host stars in binary stellar systems
$detectable$ by each one of the above mentioned techniques is then important,
since comparing the results of the observations with the predictions we can
better understand which influence close binary systems have on planet formation and evolution.
If, for example, the existence of hot-jupiters is connected to the presence of close-by 
stellar companions, we should expect to derive a higher binary frequency around stars hosting these planets, 
with respect to the frequency of binaries observed in the solar neighborhood. If, on the contrary, the 
presence of close-by stellar companions strongly prevents the existence of planets,
we should expect a lower frequency. In this paper we focus our attention on 
transit timing variations (TTVs) induced by stellar
binarity. Future contributes will account for the other techniques.


In particular here we derive the
expected $frequency$ of transiting planets in binary systems $detectable$ by TTVs. 
We define the $detection\,frequency\,(f_{det})$ as
the fraction of $transiting$ planets expected to show $detectable$
TTVs induced by stellar binarity over some fixed timescales,
when the only source of TTVs under consideration is the light travel 
time effect in binary systems.
The presence of an additional 
stellar companion around a transiting planet-host star,
should induce TTVs even if we neglect perturbing effects, because of the variable distance of
the host star with respect to the observer in the course of its orbital revolution
around the barycenter of the binary stellar system. This motion induces TTVs 
affecting the observed period of the transiting object, and consequently
the ephemerides of the transits (e.g. Irwin 1959).

Transit timing allows detenction of close stellar companions 
around more distant planet-host stars than direct imaging.
Accurate transit timing measurements are achievable also for planets discovered around faint
and distant planet-hosts, by means of a careful choice of the telescope and the 
detector (e. g. Adams et al.~2010). On the contrary, the distance of the planet-host constitutes a limit for 
direct imaging detection of stellar companions. Using VLT/NACO, and targeting solar type close-by stars ($\sim10$ pc),
a companion with a mass of $\rm 0.08\,M_{\odot}$ (then just at the limit of the brown dwarf
regime) can be detected at the 3$-\sigma$ limit at a projected separation of 0.3 arcsec 
(e.g. Schnupp et al.~2010; Eggenberger et al.~2007), which corresponds to 3$\,$AU. If, however, the star is located
at a distance $>333$ pc, direct imaging can probe only separations $>100$ AU. 
Then transit timing allows us to probe stellar multiplicity at critical separations $<100$ AU 
(as demonstrated in this work) around
more distant samples of transiting planet host-stars with respect to what can be done by direct imaging,
giving the opportunity to probe stellar multiplicity around targets located in different regions of the Galaxy.
Moreover, while direct imaging is more efficient in detecting companions in face-on orbits, transit timing
(and Doppler spectroscopy) is more efficient in the case of edge-on systems.

This paper is organized as follows: in Sect.~\ref{s:bin}, we review the known properties
of binary stellar systems in the solar neighborhood; in Sect.~\ref{s:travel_time},
we discuss the light travel time effect of transiting planets in binaries;
in Sect~\ref{s:simulations}, we describe the Monte Carlo simulations 
we did to constrain the frequency of transiting planet-host stars presenting
detectable transit timing variations induced by binarity over some fixed timescales;
in Sect.~\ref{s:results}, we discuss the results of our analysis; in Sect.~\ref{s:conclusions},
we summarize and conclude.

\section{Properties of multiple stellar systems}
\label{s:bin}

From their study of multiple stellar systems in the solar neighborhood,
Duquennoy \& Mayor~(1991) derived the following properties for binary
stellar systems with mass ratios $q>0.1$:
1) the orbital period distribution can be approximated by:

\begin{equation}
f(log\,P)\,=\,C\,exp\,\frac{-(log\,P-\overline{log\,P})^2}{2\sigma^2_{log\,P}}
\end{equation}

\noindent
where $\overline{log\,P}=4.8$, $\sigma_{log\,P}=2.3$, and $P$ is in days;
(2) binaries with periods $P>1000$ days (which is the range of periods in which we are 
interested, see below)
have an observed eccentricity distribution which tends smoothly toward
$h(e)=2e$; (3) the mass-ratio ($q=m_{2}/m_{1}$) distribution can be approximated by:

\begin{equation}
g(q)\,=\,k\,exp\,\frac{-(q-\overline{q})^2}{2\sigma^2_{q}}
\end{equation}
 
\noindent
where $\overline{q}=0.23$, $\sigma_{q}=0.42$, $k=18$ for their G-dwarf sample. 

In the following we will consider only
binaries with periods $P>5\,\cdot\,10^3$ days. This limit is due to the
need to minimize perturbing effects, which have been explicitely neglected in our analysis,
as explained in Sect.~\ref{s:travel_time}.

Using the period distribution of Duquennoy \& Mayor (1991), we have that $\sim$68$\%$ of all binary
systems are expected to have $P>5\,\cdot\,10^3$ days and, considering that the 
frequency of binary stellar systems in the solar neighborhood is 40$\%$ (as reported in Sect.~\ref{s:introduction},
excluding multiple stellar systems),
we have that the expected frequency of binaries in our period range is equal to $68\%\,\cdot\,40\%\simeq\,27\%$.

 The aim of the 
rest of this paper is
to establish how many of these systems should appear as detectable transit timing
sources over a timescale of at most 10 years since the discovery of the 
transiting planets around the primary stars of these systems.

\section{The light travel time effect}
\label{s:travel_time}

In the following we consider the case of a planet orbiting the primary star
of a binary stellar system. This configuration is 
usually called S-type orbit. We will also assume that 
the planet is revolving much closer to its parent star than
the companion star ($P_{bin}>>P_{pl}$ where $P_{bin}$
is the binary orbital period, and $P_{pl}$ is the planet orbital period).
This is a reasonable assumption given the range of binary periods 
we are considering ($P_{bin}>5\,\cdot\,10^3$ days, see above), and that
most transiting planets are typically hot-jupiters with a period of only a few days.
Under this assumption, and given also that our adopted observing window
(10 years) is smaller than the shortest period we considered, 
the perturbing effects of the secondary star on the orbit 
of the inner planet can be neglected
\footnote{In particular secular perturbations, see e.g. Kopal (1959) for a discussion
of perturbing effects of a third body on a close eclipsing binary.
}.
The three body problem can be splitted in two independent two body problems: the motion of the planet around the
host star (which can be reasonably assumed coincident with
the barycenter of the planetary system given that the mass of the planet is much smaller
than the mass of the star), and the motion of the host star
around the barycenter of the binary system.
Even neglecting perturbing effects, we expect 
transit timing variations to be present, due to the light travel time effect, as described in Sect.~\ref{s:introduction}.
The transiting planet can be used as a precise
clock unveiling the presence of the additional star in the system.

We consider a reference system with the origin in the barycenter of the binary system.
The $Z$ axis is aligned along the line of sight, the $X$ axis along the nodal 
line defined by the intersection between
the binary orbital plane and the plane of the sky, and the $Y$ axis consequently assuming
a right-handed Cartesian coordinate system. The $XY$-plane is tangent to the celestial sphere.

The distance ($z$) between the host star
and the barycenter of the binary system, projected along the observer line of sight is given by
(e. g. Kopal 1959):

\begin{equation}
z\,=\,\frac{a_1\,(1-e^2)}{1\,+\,e\,cos(f)}\,sin(\omega+f)\,sin(i),
\end{equation}

\noindent
where the orbital elements are relative to the barycentric orbit of the host star. Using the 
equation of the center of mass $a_1=m_2a/(m_1+m_2$), 
the Kepler law and dividing by the speed of light ($c$), gives:


\begin{equation}
O(t)=\frac{(G)^{1/3}\,P^{2/3}(1-e^2)\,m_2\,sin(i)}{c\,(2\pi)^{2/3}(m_1+m_2)^{2/3}}\,\frac{sin(\omega+f[t])}{1+e\,cos(f[t])},
\end{equation}

\noindent
where $f$ is the true anomaly, $e$ is the eccentricity of the orbit, and $i$ is the inclination with respect to
the plane of the sky, $m_1$ and $m_2$ are the masses of the primary and of the secondary,
$\omega$ is the argument of the pericenter, $P$ is the period, $G$ the gravitational constant,
$t$ the epoch of observation, 
and we just recall that the orbital elements are relative to the binary orbit.

Eq.~4 gives the time necessary to cover the distance projected along the 
line of sight between the host star and the barycenter of the binary system at the speed of light.
We can also say that Eq.~4 gives the difference between the 
observed ephemerides of the transiting planet once including the light time effect 
and the ephemerides obtained considering the intrinsic period of the planet
\footnote{
We are assuming that the barycentric radial velocity of the binary stellar system is $\gamma=0$.
}
.

The rate of change over time of the observed planet period with
respect to the intrinsic planet period is given by the derivative of Eq.~4 over time:


\begin{equation}
\epsilon(t)=\frac{(2\pi)^{1/3}}{c\,P^{1/3}}\,\frac{(G)^{1/3}m_2\,sin(i)}{\sqrt{1-e^2}\,(m_1+m_2)^{2/3}}\,\Big(cos(\omega+f[t])+e\,cos(\omega)\Big).
\end{equation}

\begin{figure}[!]
   \centering
   \includegraphics[width=\columnwidth]{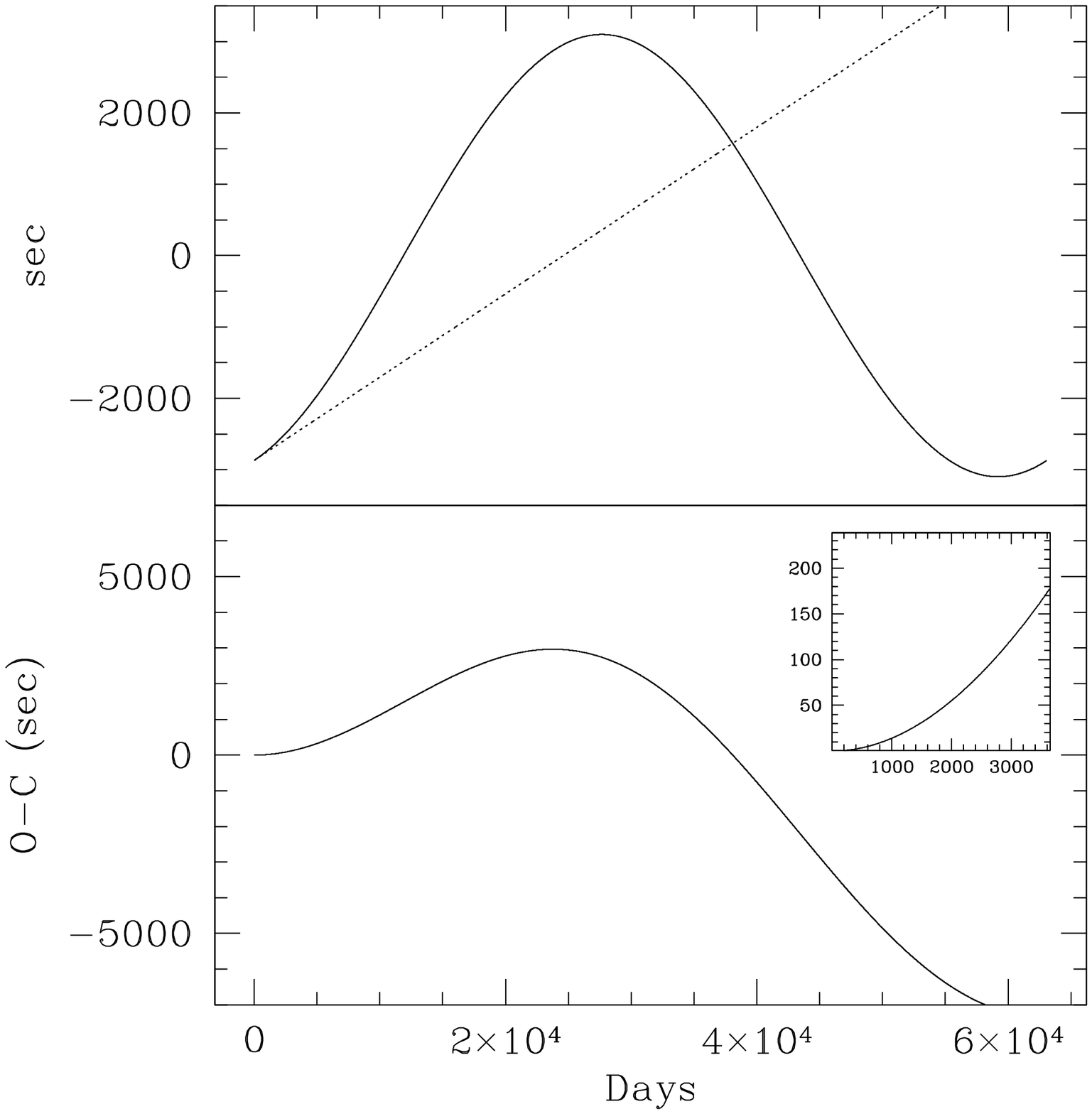}
  \includegraphics[width=\columnwidth]{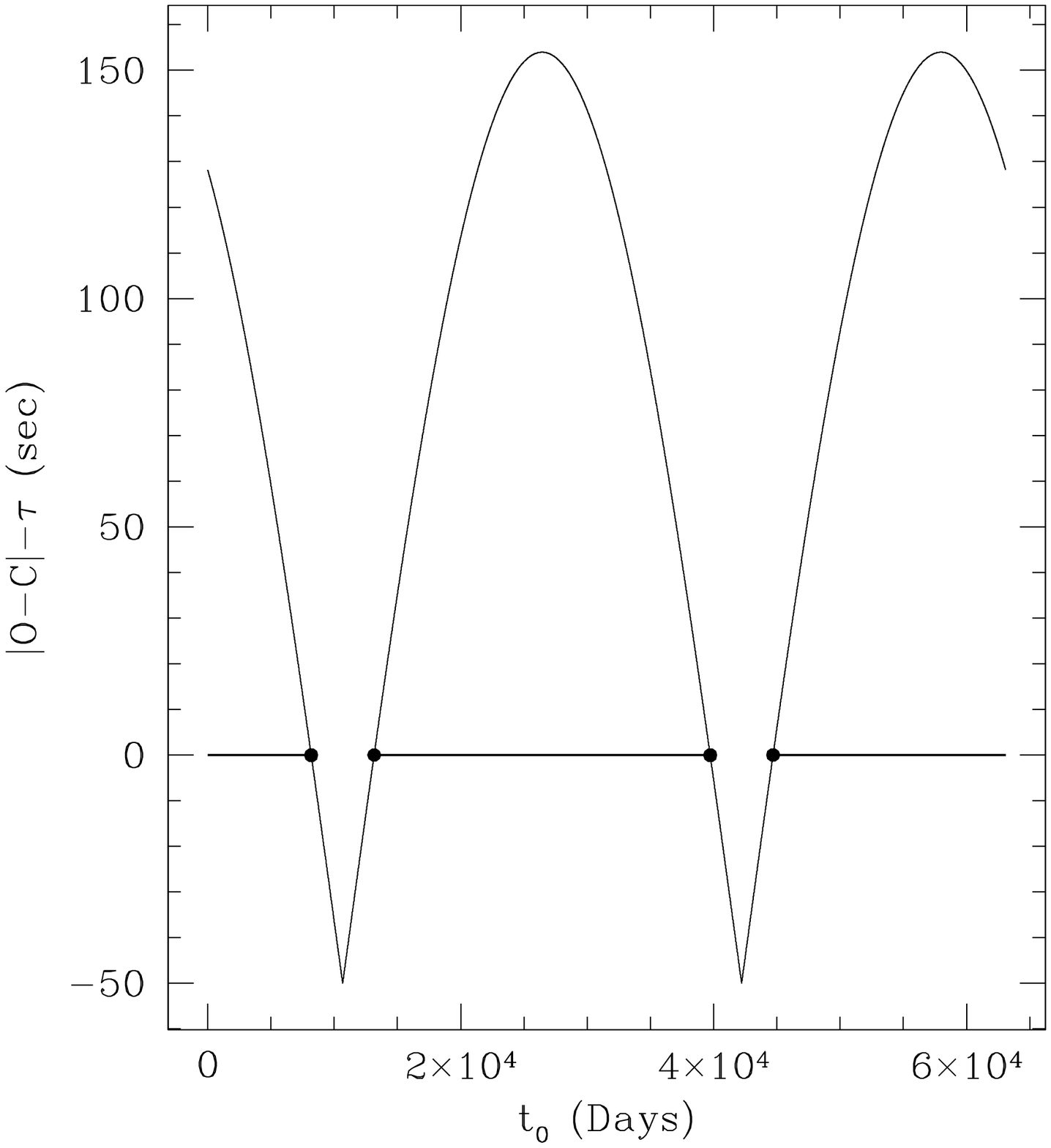}
   \caption{{\it Upper figure, upper panel:} the black solid curve represents the difference between the observed transit ephemerides
            of the planet once including the light time effect and the ephemerides obtained 
            once considering the intrinsic period of the planet
            (Eq.~4, see text). The dotted line represents the accumulation over time
            of the difference between the observed period of the planet and its intrinsic period           
            if the planet period is measured at the epoch $t_0$ coincident with the  origin of the
            x-axis, and it is then assumed constant.
            {\it Upper figure, lower panel:} the $O-C$ diagram (Eq.~6). The inner plot shows a close-up view of the $O-C$ diagram 
            during the first 10 years after the period determination. Units are the same of the large plot.
            We assumed $P_{bin}=$172.865 years, $e=0$, $q=m_2/m_1=0.23$, $\omega=0$, $i=90^{\circ}$.
            {\it Lower figure:} Graphical representation of disequation 7 in function of
            $t_0$, the epoch of the planetary period determination, for the case of
            the orbit considered in the upper figure. The horizontal
            black solid lines denote the time intervals where the condition for transit
            timing detectability is met. The value of $\tau$ is 50 sec, and the timescale
            $\Delta\,T$ is 10 years (see text for details).
            }
              \label{fig:typical_binary_curve}%
    \end{figure}

\noindent
The procedure usually adopted by observers to determine the period of a transiting
object is based on the determination
of the time interval between two or more $measured$ transits at an epoch $t_0$.
Then $\epsilon(t_0)=\epsilon_0$ can be seen as the difference between the observed and the intrinsic period at the observed time
$t_0$ due to light time effect; to avoid accumulation of this difference over time in the
ephemeris it needs to be subtracted. Fig.~\ref{fig:typical_binary_curve} (upper figure, upper panel, dotted line) shows this accumulation with
time. The difference between the observed and the predicted ephemerides ($O-C$) of successive transits
at another generic epoch
(characterized by the binary true anomaly $f$ and the epoch $t=t_0+\Delta\,t$), 
is then given by:

\begin{equation}
[O-C](t)=\,O(t)\,-\,O(t_0)\,-\,\epsilon_0\,(t\,-\,t_0).
\end{equation}

\noindent
Imposing that the $O-C$ residual is larger than a given detection threshold $\tau$ 
is equivalent to solve the following disequation:

\begin{equation}
|\,O(t_0+\Delta\,t)\,-\,O(t_0)\,-\,\epsilon_0\,(\Delta\,t)\,|\,-\,\tau\,>\,0.
\end{equation}

\noindent
In Fig.~\ref{fig:typical_binary_curve} (upper figure), we present an illustrative example. We consider a binary system where
the relevant orbital elements were fixed at the median values derived by Duquennoy \& Mayor (1991):
$P_{bin}=$172.865 years, $e=0$, $q=m_2/m_1=0.23$, $\omega=0$, $i=90^{\circ}$. We assume that a planet is orbiting
the primary star of the binary system, and that the period of the planet was measured at the epoch
$t_0$ (assumed coincident with the origin of the x-axis in Fig.~\ref{fig:typical_binary_curve}).
After five years the transit timing variation is 46 sec, and after ten years it is 178 sec, 
as shown in the inner plot of Fig.~\ref{fig:typical_binary_curve}.
The binary we have considered is one of the $typical$ 
binaries in the solar surrounding. We also observe that the $O-C$ diagram is in general 
not periodic as evident already from Fig.~\ref{fig:typical_binary_curve}.

\section{Simulations}
\label{s:simulations}

In this Section we describe the orbital simulations we did to constrain the expected frequency of 
transiting planet-host stars that should
present a detectable transit timing variation induced by stellar binarity over a 
timescale of 5 years and 10 years since the discovery of the planet around the primary star of the system.
The detectability threshold was fixed considering the results obtained by transit timing searchers
with ground based telescopes. Rabus et al.~(2009), observing the planetary system TrES-1 
with the IAC80 cm telescope obtained mean precisions of 18.5 sec, which is also in agreement with the theoretical
equation given by Doyle \& Deeg (2004). On the basis of that, and using several other literature results, 
they were already able to claim the presence of a linear trend equivalent to 48 sec (obtained from their best-fit parameter)
over the period of 4.1 yr spanned by the entire sample of the observations. 
Winn et al.~(2009) measured two transits of the giant planet WASP-4b with the Baade 6.5 m telescope
obtaining mid-transit times precisions of $\sim$6 sec. Typically it can be assumed that small or
moderate groundbased telescopes can reach precisions $<20$ sec, while large telescope may reach
better than $10$ sec precision. Given these results we considered that a TTV detection threshold equal to
$\tau=$50 sec can be reasonably applied in our analysis.

We randomly chose the period, mass ratio, and
eccentricity of the binary considering the probability distributions presented in Sect.~\ref{s:bin}. 
The mass of the primary star was fixed at $1\,M_{\odot}$, since we are considering 
binarity among typical solar type stars. The
argument of the pericenter was instead randomly chosen using a uniform probability distribution. The
inclination was either fixed to $90^{\circ}$ (to consider the case of coplanar orbits) or randomly chosen
using a uniform probability distribution. For each orbit (characterized by $P$, $e$, $q$, $\omega$, $i$)
we numerically solved the disequation (7) in function of $t_0$ (the epoch of the planetary period 
determination). We subdivided the period $P$ in 10000 equal intervals of time and evaluated 
disequation (7) at the extremes these intervals. Then we isolated the intervals in which disequation (7)
changed sign, and using the secant method imposing a
threshold for the convergence equal to 0.1 sec we obtained the roots of the correspondent equation. 
Then we determined the intervals $\Delta\,t_0$ where disequation (7) was satisfied.
Summing up together these intervals of time and dividing by the period of the binary gave the probability to observe
the requested transit timing variation for that fixed orbit over the given timescale ($\Delta\,t=t-t_0$
either 5 yr or 10 yr in our simulations) assuming to determine the period of the transiting planet 
in correspondence of a random orbital phase of the binary. In such a way we assigned to each simulated orbit
a transit timing detection probability ($P_{det}$).  In Fig.~\ref{fig:typical_binary_curve}
(lower figure), we show the 
graphical representation of disequation (7) in function of
$t_0$, the epoch of the planetary period determination, for the case of
the orbit considered in Fig.~\ref{fig:typical_binary_curve} (upper figure),
assuming a transit timing threshold $\tau=50$ sec, and a timescale of 10 years. The horizontal
black solid lines denote the time intervals where disequation (7) is satisfied.
We performed 100 runs of 10000 simulations each, calculating the mean detection probabilities
and their 1-$\sigma$ uncertainties, as reported in the next Section.

   \begin{figure}[!]
   \centering
   \includegraphics[width=\columnwidth]{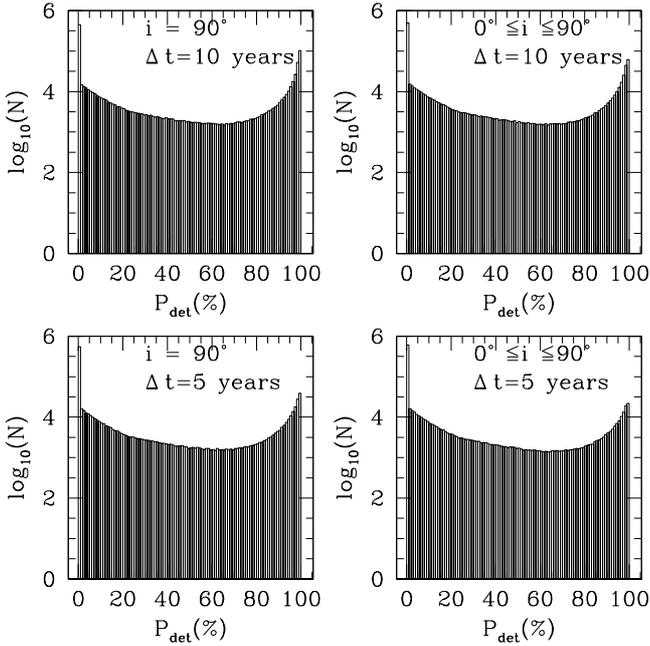}
   \caption{Distributions of the detection probabilities ($P_{det}$)
            of transit timing variations of transiting planets in binary stellar systems,
            relative to our sample of $10^6$ simulated
            binary orbits. The simulations are performed in function of different time intervals since
            the discovery of the planets
            ($\Delta\,t=5$ years or 10 years), and in function of different assumptions on the 
            inclination of the binary stellar system orbit, as indicated by the labels in the different
            panels.
    }
              \label{fig:fig_hist}%
    \end{figure}

\section{Results}
\label{s:results}

The final detection probability histograms are shown in Fig.~\ref{fig:fig_hist},
obtained from the entire sample of $10^6$ simulated orbits. While most orbits imply
null detection probabilities over the assumed timescales, in 
each one of the different situations we considered, the histograms present a probability tail extended toward
large detection probabilities. 
The number of orbits having $P_{det}>99\%$ is equal to
$3.9\%\pm0.2\%$ after 5 years since the period determination ($10.4\%\pm0.3\%$ after 10 years) 
for the case of coplanar orbits,
and is equal to $2.2\%\pm0.1\%$ after 5 years ($6.2\%\pm0.2\%$ after 10 years) for 
the case of random inclinations. 
The fact that the histograms are extended toward large probabilities is a consequence of the period
distribution and the adopted timescales. Orbits having $P_{det}>99\%$ have also 
periods smaller than $P<1.67\,\cdot\,10^5$ (considering a timescale of 10 years),
which means that transit timing can allow discovery of stellar companions up
to separations equal to $a\sim75$ AU after 10 years since the discovery of the
planet ($a\sim36$ AU after 5 years).

Then, considering the observed frequency of binaries in the solar surrounding 
with periods $P>5\,\cdot\,10^3$ days (27$\%$) and the case of
coplanar orbits, after 5 years since the discovery of a sample of transiting planets
$1.0\%\pm0.2\%$ transiting planet host-stars will have a probability $P_{det}>99\%$ to present 
$detectable$ ($>50$ sec) transit timing variations induced by stellar binarity, 
and $2.8\%\pm0.3\%$ after 10 years. 
Considering the case of random inclinations the expected frequencies ($f_{det}$) are
$0.6\%\pm0.1\%$ and $1.7\%\pm0.2\%$ after 5 and 10 years respectively.
These results are summarized in Table~\ref{tab:prob_p1_p2}. 
Our estimates can be considered a conservative lower limit, since we have 
excluded binaries with periods $P<5\,\cdot\,10^3$ days. 

\section{Conclusions}
\label{s:conclusions}

In this paper we have investigated how known transiting extrasolar planets can be used
to constrain the frequency of multiple stellar systems among planet-host stars. The presence
of a stellar companion in these systems is expected to induce transit timing variations
of the transiting planets even once perturbing effects are neglected, due to the 
orbital revolution of the primary around the barycenter of the binary stellar system. 

If the frequency of binaries among planet-host stars is the same as determined in the solar neighborhood,
after 5 years since the discovery of a sample of transiting planets
$1.0\%\pm0.2\%$ of them have a probability $>99\%$ to present a 
transit timing variations $>50$ sec induced by stellar binarity, 
and $2.8\pm0.3\%$ after 10 years if the planetary and binary orbits are coplanar. Considering the case 
of random inclinations the probabilities are $0.6\%\pm0.1\%$ and $1.7\%\pm0.2\%$ after 5 and 10 years respectively.
Our results have been obtained assuming a binary period $P>5\,\cdot\,10^3$ ($a\gtrsim6$  AU). Moreover, we derived that  
we can expect to discover stellar companions of transiting planets host stars up to a maximum separations
$a\sim75$ AU after 10 years since the discovery of a planet ($a\sim36$ AU after 5 years).

A final comment is necessary to mention that TTVs may have several different origins among which
perturbing effects caused by additional planets or moons, secular precession due to general relativity,
stellar proper motion, the Appelgate effect
(Agol et al. 2005; Miralda-Escud\'e 2002; Nesvorn\'y 2009; Heyl \& Gladman 2007; 
Ford \& Holman 2007; Simon 2007; Kipping 2009a; Kipping 2009b; P\'al \& Kocsis 2008;
Rafikov 2009; Watson 2010),
and binarity of the host is one of them. However, transit timing 
variations induced by binarity are expected to produce
long-term trends, and in particular they should be associated also with radial velocity
drifts of the host star. Transit timing searchers should also
follow-up spectroscopically their targets, since a transit
timing variation associated with a radial velocity variation will be very likely the signature of binarity.

\begin{table}
\caption{Expected frequency ($f_{det}$) of binary stellar systems detectable
by means of transit timing variations in function of the timescale ($\Delta\,T$) since the discovery of the planet,
and of different assumptions on the inclination of the binary stellar systems. 
}
\label{tab:prob_p1_p2}
\centering
\begin{tabular}{c c c c}
\hline
\hline
$i=90^{\circ}$ & $\Delta\,T$ & $f_{det}$ \\
 & (yr) & ($\%$) \\
\hline
 & 5 & $1.0\pm0.2$ \\
 & 10 & $2.8\pm0.3$ \\
\hline
\hline
$0^{\circ}\le\,i\le\,90^{\circ}$ & $\Delta\,T$ & $f_{det}$ \\
 & (yr) & ($\%$) \\
\hline
 & 5 & $0.6\pm0.1$ \\
 & 10 & $1.7\pm0.2$ \\
\hline
\end{tabular}
\end{table}

\acknowledgements{
The author is grateful to Dr. Daniel Fabrycky, Dr. Francesco Marzari,
and Dr. Silvano Desidera for interesting comments and discussions about planets in binary
stellar systems, and to the anonymous referee for his/her useful comments and
suggestions.
}

\end{document}